\documentclass[reprint,
 amsmath,
 amssymb,
 floatfix]{revtex4-1}

\usepackage{amsmath,amssymb,amsfonts,dsfont}
\usepackage{mathcomp}
\usepackage[normalem]{ulem}
\usepackage[usenames, dvipsnames]{color}
\usepackage{graphicx}
\usepackage{dcolumn}
\usepackage[caption=false]{subfig}
\usepackage[unicode=true, breaklinks=false, pdfborder={0 0 1}, backref=false, colorlinks=true, linkcolor=blue, urlcolor=blue, citecolor=blue,bookmarksopen=true]{hyperref}
\hypersetup{colorlinks=true} 
\usepackage{units}

\newcommand{\tr}{\operatorname{tr}}

\pdfoutput=1

\begin{document}

\title{Synchronisation phase as an indicator of persistent quantum correlations between subsystems}

\author{Stefan Siwiak-Jaszek}
\email{stefan.siwiak-jaszek.11@ucl.ac.uk}
 
\author{Thao P. Le}

\author{Alexandra Olaya-Castro}%
 \email{a.olaya@ucl.ac.uk}
 
\affiliation{Department of Physics and Astronomy, University College London, London WC1E 6BT, United Kingdom}%

\date{\today}

\begin{abstract}
Spontaneous synchronisation is a collective phenomenon that can occur in both dynamical classical and quantum systems. 
Here, we analyse the spontaneous synchronisation dynamics of vibrations assisting energy transfer in a bio-inspired system. We find the emergence of a constant non-zero `synchronisation phase' between synchronised vibrational displacements as the natural frequencies of the oscillators are detuned.
This phase difference arises from the asymmetric participation of local modes in the long-lived synchronised state. Furthermore, we investigate the relationships between the synchronisation phase, detuning and the degree of quantum correlations between the synchronising subsystems and find that the synchronisation phase captures how quantum correlations persistently exceed classical correlations during the dynamics. We show that our analysis applies to a variety of spontaneously synchronising open quantum systems. Our work therefore opens up a promising avenue to investigate non-trivial quantum phenomena in complex biomolecular and nano-scale chemical systems. 

\end{abstract}

\keywords{synchronisation; quantum; quantum synchronisation; bio-inspired; quantum coherence; exciton; photosynthesis; synchronisation phase; quantum correlations; quantum discord}
\maketitle

\section{\label{sec:intro}Introduction}
When two oscillators interact, they can undergo \emph{synchronisation} where their frequencies align \cite{Pikovsky2001}, usually with a nonzero phase difference. Synchronisation is a well studied classical phenomenon that appears recurrently throughout the natural world \cite{Buck1966, Nitsan2016}. 
More recently it has been explored for physical systems in the quantum domain \cite{Lorch2017,Giorgi2016,Qiu2015,Hush2015,Walter2014a,Lee2013, Galve2016,Eneriz2019}.
Oscillating quantum systems with a stable limit cycle can synchronise in the steady state to an external driving field \cite{Walter2014a} or to another quantum system \cite{Lee2013}. The latter is often referred to as spontaneous synchronisation \cite{Cabot2019}. In the absence of any external drive, quantum systems can also undergo \emph{transient} spontaneous quantum synchronisation during the early time dynamics prior to relaxation to the ground state \cite{Giorgi2012, Giorgi2013,Benedetti2016,Manzano2013,Militello2017,Cabot2019}. The underlying mechanisms leading to this transient behaviour are effective collective dissipation processes \cite{Giorgi2013} or combined local dissipation and system-system interactions \cite{Manzano2013, Cabot2019}. Transient spontaneous quantum synchronisation is particularly interesting in bio-inspired quantum settings as it allows us to understand the possible roles of quantum coherence in such systems \cite{Siwiak-Jaszek2019}. Furthermore, transient spontaneous synchronisation has been linked to the appearance of quantum correlations \cite{Giorgi2013, Manzano2013, Cabot2019}, suggesting they may be necessary for synchronisation to occur \citep{Cabot2019}.
However, research on transient spontaneous quantum synchronisation rarely focuses on the nature of \emph{phase} itself between synchronised systems, despite its classical relevance.

Classically, the `robustness' of  synchronisation refers to the ability of oscillators to lock in phase despite their different natural frequencies: within a certain range of detunings $\Delta \nu$, the oscillators are able to lock into a new resultant frequency that lies between the original frequencies.  The equation of motion for the phase difference between two weakly coupled classical oscillators to first order is \cite{Pikovsky2001}:
\begin{equation}
    \frac{d\Delta \phi(t)}{dt} = -\Delta \nu + \epsilon f\left(\Delta \phi(t)\right),
    \label{eqn: classical phase equation}
\end{equation}
where $\Delta \phi(t)$ is the phase difference between oscillations, $\epsilon$ is a constant proportional to their coupling strength, and $f(\cdot)$ is a time-dependent periodic function. Synchronisation can only occur if the detuning lies between the min and max extrema of $f\left(\Delta \phi(t)\right)$:
\begin{equation}
    \min_\phi \epsilon f\left(\Delta \phi(t)\right) < \Delta \nu < \max_\phi \epsilon f\left(\Delta \phi(t)\right).
\end{equation}
Within this detuning region, there is at least one stationary solution to Eq.~(\ref{eqn: classical phase equation}) corresponding to synchronisation. In most cases this synchronisation occurs with a non-zero phase.  We will refer to this phase difference as the \emph{synchronisation phase}. As the detuning $\Delta \nu$ increases, the synchronisation degrades and eventually disappears.

Quantum mechanical oscillators typically exhibit analogous behaviour regarding detuning and synchronisation to their classical counterparts as described above \cite{Giorgi2013}. Analytic equations similar to Eq.~(\ref{eqn: classical phase equation}) have been derived in non-linear many-body quantum systems \cite{Witthaut2017} and  exciton-polariton condensates \cite{Wouters2008}. However, in contrast to classical oscillators, detuning can \emph{enhance} steady-state synchronisation of Van der Pol oscillators operating in the deep quantum regime  \cite{Lorch2017}. Furthermore, synchronisation phase can emerge for identical quantum harmonic oscillators interacting with a common two-level system \cite{Militello2017}.

Here, we employ a modified version of the Pearson correlation factor as a measure of synchronisation \cite{Siwiak-Jaszek2019} to investigate relationships between synchronisation phase, detuning and quantum correlations in open quantum systems exhibiting transient spontaneous synchronisation. In particular, we consider the synchronisation of vibrational displacements in a bio-inspired system featuring exciton-vibration interactions as observed in some photosynthetic complexes (see for example \cite{Richards2012, Dean2016, Romero2014a, Fuller2014}). It has been argued that these interactions are a mechanism for sustaining coherent processes in such molecular complexes \cite{Kolli2012, Tiwari2012, Chin2013, Huelga2013, OReilly2014, Romero2014a, Fuller2014, Novelli2015}. Thus, investigating this bio-inspired vibronic system can shed light on the synchronisation and quantum correlation processes in biophysical systems operating at the interface between the classical and quantum regimes.

We study the dynamics of this bio-inspired system assuming a detuning between the natural frequencies of the vibrations of interest and show the emergence of a constant phase difference in their spontaneous synchronised states. We discuss how the synchronisation phase occurs due to the break of symmetry both in the Hamiltonian and in the dynamics with respect to mode exchange. In order to gain further insight into the information captured by the synchronisation phase, we investigate the quantum correlations between the subsystems of interest as measured by the quantum discord \cite{Henderson2001, Ollivier2002}. We show that for spontaneous quantum synchronisation to emerge, the quantum discord must be greater than classical information at all times. We further show that our synchronisation measure is able to capture a change in quantum discord between subsystems  as  a  function  of  frequency detuning. By extending our analysis to the model considered in Ref.~\cite{Militello2017}, we note that the identified relationships between synchronisation, detuning and quantum correlations hold in a variety of open quantum system scenarios.

This paper is organised as follows: in Section \ref{sec:methods}, we describe the bio-inspired vibronic system, its open system evolution, and the measure for transient spontaneous synchronisation we employ. In Section \ref{sec:Results}, we present the effects of detuning on synchronisation in the dimer. In Section \ref{sec:quantum correlations}, we study the relationship between quantum correlations and synchronisation phase. In Section \ref{sec: militello et al case study}, we contrast our results with that of Ref. \cite{Militello2017} in which synchronisation phase occurs without dependence on detuning. Finally, we conclude in Section \ref{sec:discussion_and_conclusion}.

Note that for the rest of the paper, we are focused entirely on transient spontaneous quantum synchronisation, thus we will typically refer to it simply as synchronisation.

\section{\label{sec:methods}Modelling transient spontaneous synchronisation in exciton-vibration dimers}

In Section \ref{sec:exciton-vibration dimer} we introduce the Hamiltonian for the exciton-vibration dimer model. In Section \ref{sec:Open Systems Model} we describe the Markovian master equation for the open quantum system dynamics of the exciton-vibration dimer and our numerical methods. In Section \ref{sec:synch measure} we describe the measure used to quantify transient spontaneous quantum synchronisation and discuss its limitations.

\subsection{The exciton-vibration dimer with detuning} \label{sec:exciton-vibration dimer}

\begin{figure}
    \centering
    \includegraphics[width=\columnwidth]{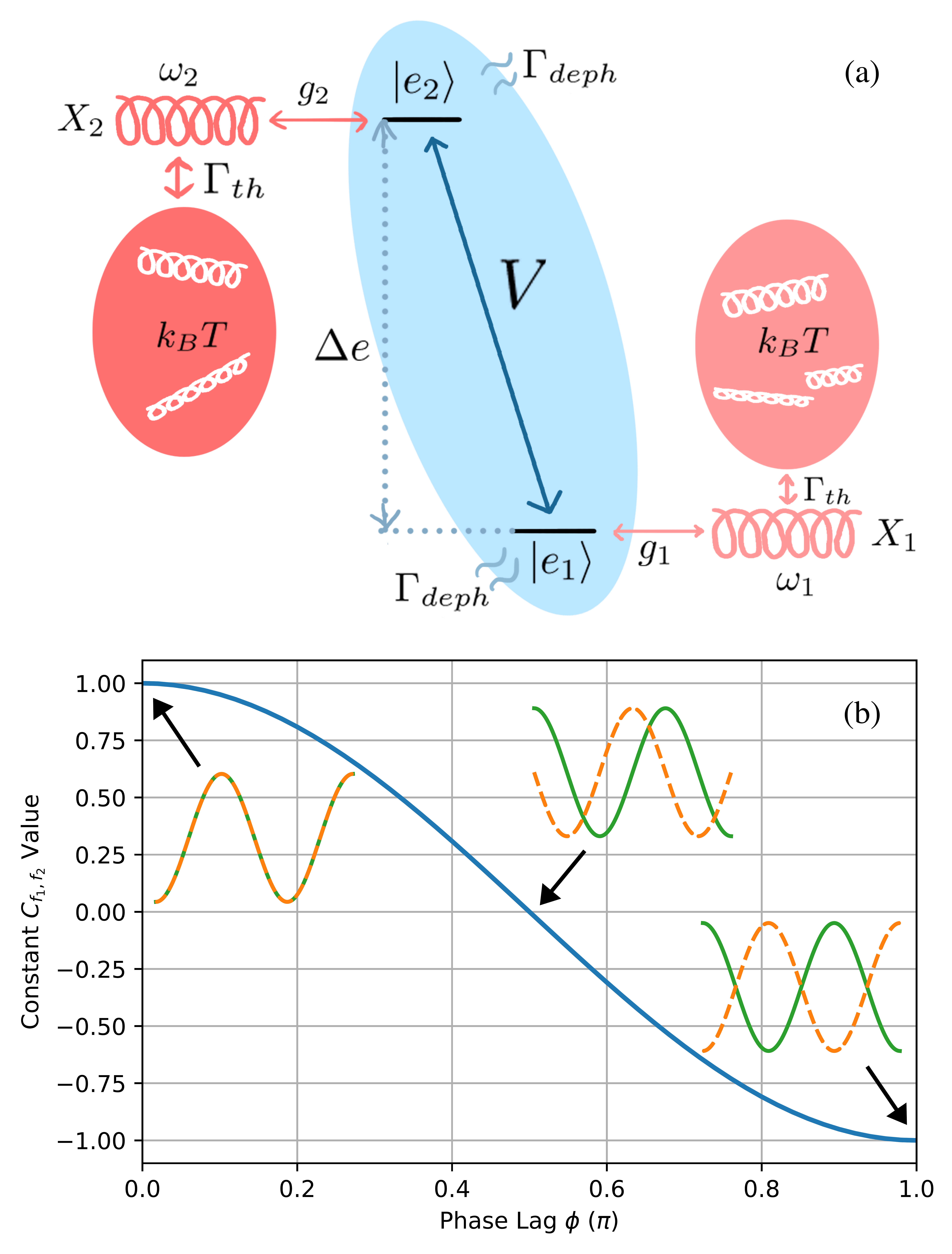}\hfill
    \caption{(a) Schematic diagram of the exciton-vibration dimer. Two chromophores (subscripts $i=1,2$) with single excited states $|e_i\rangle$ interact via dipole-dipole coupling of strength $V$. Each electronic state is coupled linearly with strength $g_i$ to a harmonic mode of energy $\omega_i$. The electronic subsystem (central blue oval) experiences pure dephasing of rate $\Gamma_{deph}$, and each mode dissipates into separate thermal baths (red ovals) of temperature $k_BT$ at rate $\Gamma_{th}$. (b) Value of the transient spontaneous quantum synchronisation measure $C_{f_1,f_2}(\Delta t ~= ~2 \pi / a) = \cos\phi$  [Eq.~(\ref{eqn:synchro_for_specific_f1_f2})] for two identical sinusoids $f_1, f_2$ as a function of their phase difference $\phi$, where $f_1 = \sin (at)$, $f_2=\sin(at+\phi)$.} \label{fig:model_C_phase_lag}
\end{figure}

The \emph{exciton-vibration dimer} is a prototype light-harvesting unit formed by a pair of chromophores whose local electronic excitations interact with quasi-coherent vibrational modes \cite{Kolli2012,OReilly2014}. We have previously explored synchronisation using this model in Ref.~\cite{Siwiak-Jaszek2019} and here we introduce a modified version to allow us to explore detuning and synchronisation.

The exciton-vibration dimer is composed of two chromophores and two harmonic modes. The chromophores have single excited states $|e_{i=1,2}\rangle$ with energies $e_{i=1,2}$, and they interact via dipole-dipole coupling of strength $V$. Each chromophore is then locally coupled to a quantised intramolecular mode of energy $\omega_{i=1,2}$ with strength $g_{i=1,2}$. Fig.~\ref{fig:model_C_phase_lag}(a) contains a diagram of the dimer model. In this detuned situation, we must account properly for the different reorganisation energy contributions to the Hamiltonian \cite{MayKuhnBook2011,Renger2004,Stones2017}, in order to accurately account for the effects of different frequency modes on the dynamics of the exciton-vibration dimer. For clarity, in the following we explicitly derive the dimer Hamiltonian relevant for our situation.

The Hamiltonian for a two-level electronic system, with each site locally and linearly couples to a vibrational mode \cite{Renger2004, Schatz2002} can be written as follows:
\begin{equation}
    \begin{split}
    H = &  \left(e_1+\frac{1}{2}\omega_1^2d_{1}^2\right)
    | e_1\rangle\langle  e_1| \\
    &+ \left(e_2 + \frac{1}{2}\omega_2^2d_{2}^2\right)
    | e_2\rangle\langle  e_2| \\
    & + \frac{1}{2}\omega_1^2\hat{x}_1^2 + \frac{1}{2}\hat{p}_1^2 + \frac{1}{2}\omega_2^2\hat{x}_2^2 + \frac{1}{2}\hat{p}_2^2 \\
    & - \omega_1^2\hat{x}_1d_{1} | e_1\rangle\langle  e_1| - \omega_2^2\hat{x}_2d_{2} | e_2\rangle\langle  e_2| \\
    & + V\left(| e_2\rangle\langle e_1| + | e_1\rangle\langle  e_2|\right),
    \end{split}\label{eq:initial_H}
\end{equation}
where $e_i$ ($i=1,2$) are the energies of the bare electronic states, $\hat{x}_i$ and $\hat{p}_i$ are the position and momentum operators of mode $i$ coupled to site $i$, and $d_{i}$ is the displacement of the equilibrium position of mode $i$ due to electronic state $|e_i\rangle$. This displacement is effectively the site-mode coupling strength.

Now we can define the reorganisation energy $\lambda_i = \frac{1}{2}\omega_i^2d_{i}^2 = \omega_iS_i$ contribution of the mode to the site energy, where $S_i$ is the Huang-Rhys factor which is experimentally observable through measurements of the Stokes shift \cite{MayKuhnBook2011}. If $\omega_1=\omega_2$ and $d_{1}=d_{2}$, the site energies are both shifted by the same amount and the reorganisation energy has no affect on dynamics. However, as we are interested in $\omega_1 \neq \omega_2$, the reorganisation energy contributions to site energies cannot be discarded.

We can write the position and momentum operators in terms of the creation and annihilation operators: $\hat{x}_i = \frac{1}{\sqrt{2}} \sqrt{\frac{1}{\omega_i}} \left(b^\dag_i+b_i\right)$ and $\hat{p}_i = \frac{i}{\sqrt{2}} \sqrt{\omega_i} \left(b_i^\dag-b_i\right)$, where $b_{i=1,2}^\dag(b_{i=1,2})$ are the creation (annihilation) operators for the modes. Substituting these into the previous expression Eq.~(\ref{eq:initial_H}), we obtain:
\begin{equation}
    \begin{split}
    H = & \left(e_1+\omega_1S_1\right)
    | e_1\rangle\langle  e_1| + \left(e_2 + \omega_2S_2\right)
    | e_2\rangle\langle  e_2| \\
    & + \omega_1\left(b_1^\dag b_1+\frac{1}{2}\right) + \omega_2\left(b_2^\dag b_2+\frac{1}{2}\right) \\
    & - \omega_1\sqrt[]{S_1}\left(b_1^\dag+b_1\right)| e_1\rangle\langle  e_1| \\
    & -  \omega_2\sqrt[]{S_2}\left(b_2^\dag+b_2\right)| e_2\rangle\langle  e_2| \\
    & + V\left(| e_2\rangle\langle e_1| + | e_1\rangle\langle  e_2|\right).
    \end{split}
\end{equation}
We then rotate into a new frequency-dependent exciton basis with matrix $U(\Tilde{\theta}(\omega_1,\omega_2)) $\cite{MayKuhnBook2011},
\begin{equation}
    U = \begin{pmatrix}\cos\Tilde{\theta} & \sin\Tilde{\theta} \\ -\sin\Tilde{\theta} & \cos\Tilde{\theta}\end{pmatrix},
\end{equation}
where $\Tilde{\theta}(\omega_1,\omega_2) = \frac{1}{2}\arctan\left(\frac{2|V|}{(e_2 + \omega_2S_2) - (e_1+\omega_1S_1)}\right)$ is the mixing angle and can be used as a measure of electronic delocalisation and hence the exciton size. We then shift the ground state energy of mode 1 by $\frac{\omega_1}{2}$ and mode 2 by $\frac{\omega_2}{2}$.
The final total Hamiltonian for our exciton-vibration dimer is then:
\begin{equation}
    \begin{split}
    H = & + E_1(\omega_1,\omega_2)|E_1\rangle\langle E_1| + E_2(\omega_1,\omega_2)|E_2\rangle\langle E_2| \\ & + \omega_1 b^\dag_1b_1 + \omega_2 b^\dag_2b_2 \\ & + \omega_1\sqrt[]{S_1} \Tilde{\Theta}_1 X_1 + \omega_2\sqrt[]{S_2} \Tilde{\Theta}_2 X_2, 
    \end{split} \label{System Hamiltonian}
\end{equation}
where each line corresponds to the exciton Hamiltonian, vibrational Hamiltonian, and exciton-vibration interaction Hamiltonian respectively. We have defined $\Tilde{\Theta} = U\left(\Tilde{\theta}(\omega_1,\omega_2)\right)|e_i\rangle \langle e_i|U^\dag\left(\Tilde{\theta}(\omega_1,\omega_2)\right)$, and the position operator for each mode are $X_{i=1,2} = b_i+b_i^\dag$. The excitons $|E_{d=1,2}\rangle$ have eigenenergies that are mode-frequency dependent:
\begin{equation}
    \begin{split}
    E_{i=1,2}(\omega_1,\omega_2) =  &\frac{1}{2}\big[
    (e_1 + \omega_1S_1)+(e_2 +\omega_2S_2) \\
     & \quad + (-1)^i \, \sqrt[]{\Delta e^2(\omega_1,\omega_2) + 4V^2}\big],
    \end{split}
\end{equation}
where $\Delta e (\omega_1,\omega_2) = (e_2 + \omega_2S_2) - (e_1+\omega_1S_1)$. 

The eigenstates of the total Hamiltonian $H$ are exciton-vibrational which we can write in the local basis as:
\begin{equation}
    \begin{split}
    |\psi_j\rangle = & \sum_{d=1,2}\alpha_d|E_d\rangle \otimes \sum^{M}_{n_1=1}\beta_{n_1} |n_1\rangle\otimes  \sum^{M}_{n_2=1}\gamma_{n_2} |n_2\rangle \\ = & \sum_{d,n_1,n_2}c(d,n_1,n_2)|E_d,n_1,n_2\rangle,
    \end{split} \label{eq:eigenstates_for_H}
\end{equation}
where eigenstates $|\psi_j\rangle$ are labelled in ascending energy, and  $|n_1\rangle \otimes |n_2\rangle$ where $n$ are the Fock state numbers and subscripts indicate the mode subspace. To obtain convergent dynamics we account for a maximum occupation $M=8$ in each mode. 

Note that by including the reorganisation energy, large detunings will have a significant effect on the coherent dynamics of the dimer. In the following, the detunings and associated reorganisation energies considered are small enough for the overall features of the excitation dynamics to be subtle yet they do affect the correlations involved in the synchronisation process.

In this paper, we consider the regime of weak electronic coupling where $\Delta E\approx\omega>g>V$ which is characteristic of chromophore pairs present in a variety of light-harvesting proteins \cite{Womick2011, Richards2012, Kolli2012,Viani2014,Doust2004,Novoderezhkin2010,Novelli2015,Collini2010}.
We investigate the synchronisation of oscillations in the expectation value of the position operators, $\langle X_i \rangle$, which we refer to as local mode displacements. In the regime we consider, excitons are not fully delocalised and excitonic energies are close to the local energies. The vibronic eigenstates are of a quasi-localised nature and an analysis of their dynamics is therefore useful for understanding the synchronisation of local mode displacements \cite{Siwiak-Jaszek2019}.

\subsection{Open quantum system model}\label{sec:Open Systems Model}

Let $\rho(t)$ be the density matrix of the exciton-vibration dimer.  In addition to its Hamiltonian from Eq.~(\ref{System Hamiltonian}), the dimer also undergoes relaxation caused by its surrounding environment. We consider this process to be Markovian and of Lindblad form,
\begin{equation} \label{eq: master}
    \dot{\rho}(t) = -i[H,\rho(t)] + D_{deph}[\rho(t)] + D_{th}[\rho(t)],
\end{equation}
where $D_{deph}$ and $D_{th}$ are Lindblad superoperators of the form:
\begin{equation}
D_{\nu}[\rho] = \Gamma_{\nu}\left(O_{\nu} \rho O^\dag_{\nu} - \frac{1}{2}\rho O_{\nu}^\dag O_{\nu} - \frac{1}{2}O^\dag_{\nu} O_{\nu}\rho\right),
\label{eq:lindblad}
\end{equation}
for various operators  $O_\nu$  at rates $\Gamma_{\nu}$.

We assume that each electronic system undergoes local pure dephasing \cite{Haken1973,Breuer2002} with operators $|e_1\rangle\langle e_1|$ and $|e_2\rangle\langle e_2|$ at equal rates of $\Gamma_{deph}=\left[ \unit[0.1]{ps}\right]^{-1}$ such that exciton coherence decays in approximately $\unit[0.5]{ps}$, matching the experimental evidence of algal photosynthetic proteins. \cite{Richards2012}.

We assume that each mode undergoes relaxation \cite{Breuer2002} due to thermal reservoirs at temperature $\unit[298]{K}$ ($\unit[207.1]{cm^{-1}}$). This corresponds to transition Lindblad operators $b_1$ and $b_2$ at rate $\Gamma_{th}(1+B)$, and $b^\dag_1$ and $b^\dag_2$ at rate $\Gamma_{th}B$. Here $B = (e^{\frac{\omega}{k_BT}}-1)^{-1}$ is the mean number of quanta in a thermally occupied mode of frequency $\omega$ and $\Gamma_{th}=\left[ \unit[1]{ps}\right]^{-1}$ is the rate at which modes equilibrate. Table \ref{table: numerical parameters} summarises the various parameters.

\begin{table}
    \begin{center}
    \begin{ruledtabular}
    \begin{tabular}{c|c|c|c|c|c|c}
    $\Delta e$&$V$&$\omega$&$g$&$k_BT$&$\Gamma_{th}$&$\Gamma_{deph}$\\
    \hline
    1042&92&1111&267.1&207.1& $\left[ \unit[1]{ps}\right]^{-1}$ &  $\left[ \unit[0.1]{ps}\right]^{-1}$
    \end{tabular}
    \end{ruledtabular}
    \end{center}
    \caption{Parameters used for numerical calculations representing the central dimer in the cryptophyte photosynthetic antennae PE545 \cite{Novoderezhkin2010,Kolli2012}. All units in spectroscopic wavenumbers $\unit{cm^{-1}}$ except for the rightmost two columns which are specified in table.}
    \label{table: numerical parameters}
\end{table}

To numerically simulate the model, we linearise the master equation into an ordinary differential equation
\begin{equation}
    |\dot{\rho}(t)\rangle \mkern-3mu \rangle = \mathcal{L}|\rho(t) \rangle \mkern-3mu \rangle,
\end{equation}
where $\mathcal{L}$ is the Liouvillian superoperator and $|\rho(t) \rangle \mkern-3mu \rangle$ are flattened density matrices, which can be solved with standard algorithms.

Our initial state is
\begin{equation}
    \rho(0) = |E_2\rangle\langle E_2|\otimes \rho^{th}_1\otimes\rho^{th}_2,
    \label{eqn: initial state}
\end{equation}
where the electronic system starts in the higher energy excitonic state $|E_2\rangle$, and both intramolecular modes are initially in thermal equilibrium with their respective baths: $\rho_{i}^{th}=\sum_{n_i}P_{n_i}|n_i\rangle\langle n_i|$ and $P_{n_i} = \big(1-e^{\frac{-\omega}{k_BT}}\big) e^{\frac{-n_i \omega}{k_BT}}$.

\subsection{Measuring transient spontaneous synchronisation}\label{sec:synch measure}

Typically, the Pearson correlation factor is used in the quantum synchronisation literature as it gives clear indication of synchronisation and shows some information about the timescales involved. For any two time dependent functions $f_1(t)$ and $f_2(t)$, the Pearson correlation coefficient is defined as follows \cite{Galve2016}:
\begin{equation}
C_{f_1,f_2}\left(t|\Delta t\right) = \frac{\int_{t}^{t+\Delta t} \delta f_1 \delta f_2 dt}{\left(\int_{t}^{t+\Delta t}\delta f_1^2 dt \int_{t}^{t+\Delta t}\delta f_2^2 dt\right)^{1/2}},
\label{eqn: correlationcoeff}
\end{equation}
where $\delta f = f - \bar{f}$, $\bar{f} = \frac{1}{\Delta t} \int_{t}^{t+\Delta t} f\left(t'\right) dt'$ is a time average and $\Delta t$ is the averaging window.

However, the usual form of the Pearson correlation factor does not give sufficient information about the early-time transient dynamics before synchronisation is reached, instead only indicating when it has occurred. Here we present a modification to the Pearson correlation factor that allows it to be used as a continuous measure of phase difference and therefore reveal information about the early times and the emergence of synchronisation. This was first presented in Ref. \cite{Siwiak-Jaszek2019}. To do so, we make a particular choice of $\Delta t$, such that our measure of synchronisation is:
\begin{equation}
C_{\langle X_1 \rangle,\langle X_2 \rangle}\left(t \bigg|\Delta t = \dfrac{2 \pi}{\omega_1}\right),
\label{eqn: correlationcoeff_synchronisation_measure}
\end{equation}
where $\omega_1$ is the frequency of oscillation of the first chromophore, and $\langle X_i \rangle$ are the expectation values of the position operator for each mode. We explain how it works in the remainder of this section.

The Pearson correlation factor returns a value of $1$ for positive spontaneous synchronisation (in-phase), $-1$ for negative synchronisation ($\pi$ out of phase) and $0$ for asynchrony \cite{Galve2016}. If we chose $\Delta t$  as close as possible to time period $T$ of the dominant frequency in $f_{1}$, then the correlation $C_{f_1,f_2}\left(t|\Delta t=T\right)$ becomes a continuous measure of phase difference between the two oscillating signals. This function returns a continuous value in the range of $-1$ to $1$ corresponding to a phase shift of $\pi$ to $0$ respectively between the functions $f_1$ and $f_2$.

We derive this relation analytically for two example sinusoids in the following. Consider $f_1 = \sin\omega t$ and $f_2 = \sin(\omega t +\phi)$ with identical frequencies and amplitudes that lie within the same range. With the choice of $\Delta t = \frac{2\pi}{\omega}$, we find that both time averages $\bar{f_1}$ and $\bar{f_2}$ are zero:
\begin{align}
    \bar{f_1} &= \frac{\omega}{2\pi} \int^{t^\prime +\frac{2\pi}{\omega}} _{t^\prime} \sin \omega t dt = 0,\\
    \bar{f_2} & = \frac{\omega}{2\pi} \int^{t^\prime +\frac{2\pi}{\omega}} _{t^\prime} \sin(\omega t +\phi) dt  = 0.
\end{align}

Note that in general, the functions $f_1$ and $f_2$ can have different amplitudes or be displaced from zero. Hence, in general their time averages $\bar{f}_1$ and $\bar{f}_2$ are not zero. The shifted functions $\delta f_i = f_i - \bar{f_i}$ act to subtract the average value of $f_i$ and center any oscillations around zero. This emphasises the fluctuations around the mean and allows more accurate measurement of phase. For our example  $f_1$ and $f_2$, we have $\delta f_i = f_i$. The integral of their product is the main measure of synchronisation:
\begin{equation}
\begin{split}
    \int^{t^\prime +\frac{2\pi}{\omega}} _{t^\prime} \delta f_1 \delta f_2 dt  = &  \int^{t^\prime +\frac{2\pi}{\omega}} _{t^\prime} \sin(\omega t) \sin(\omega t +\phi) dt \\
    = & \frac{\pi \cos\phi}{\omega}.
\end{split}
\label{eqn: synchronisation product integral}
\end{equation}
This shows that for a sliding window of one time period $T= \frac{2\pi}{\omega}$ and two perfect sinusoids, the time dependence disappears. For any value of $t$, the measure returns a constant value that depends only on the phase difference $\phi$ between the signals. 

The denominator in Eq.~(\ref{eqn: correlationcoeff}) normalises the measure to the limits of $1$ and $-1$. For our example oscillations it takes value $\left(\int_{t}^{t+\Delta t}\delta f_1^2 dt \int_{t}^{t+\Delta t}\delta f_2^2 dt\right)^{1/2} = \frac{\pi}{\omega}$. Together, this finally results in the synchronisation measure:
\begin{equation}
    C_{f_1,f_2}\left(\Delta t = T\right) = \cos\phi. \label{eqn:synchro_for_specific_f1_f2}
\end{equation}

\begin{figure}
    \centering
    \includegraphics[width=\columnwidth]{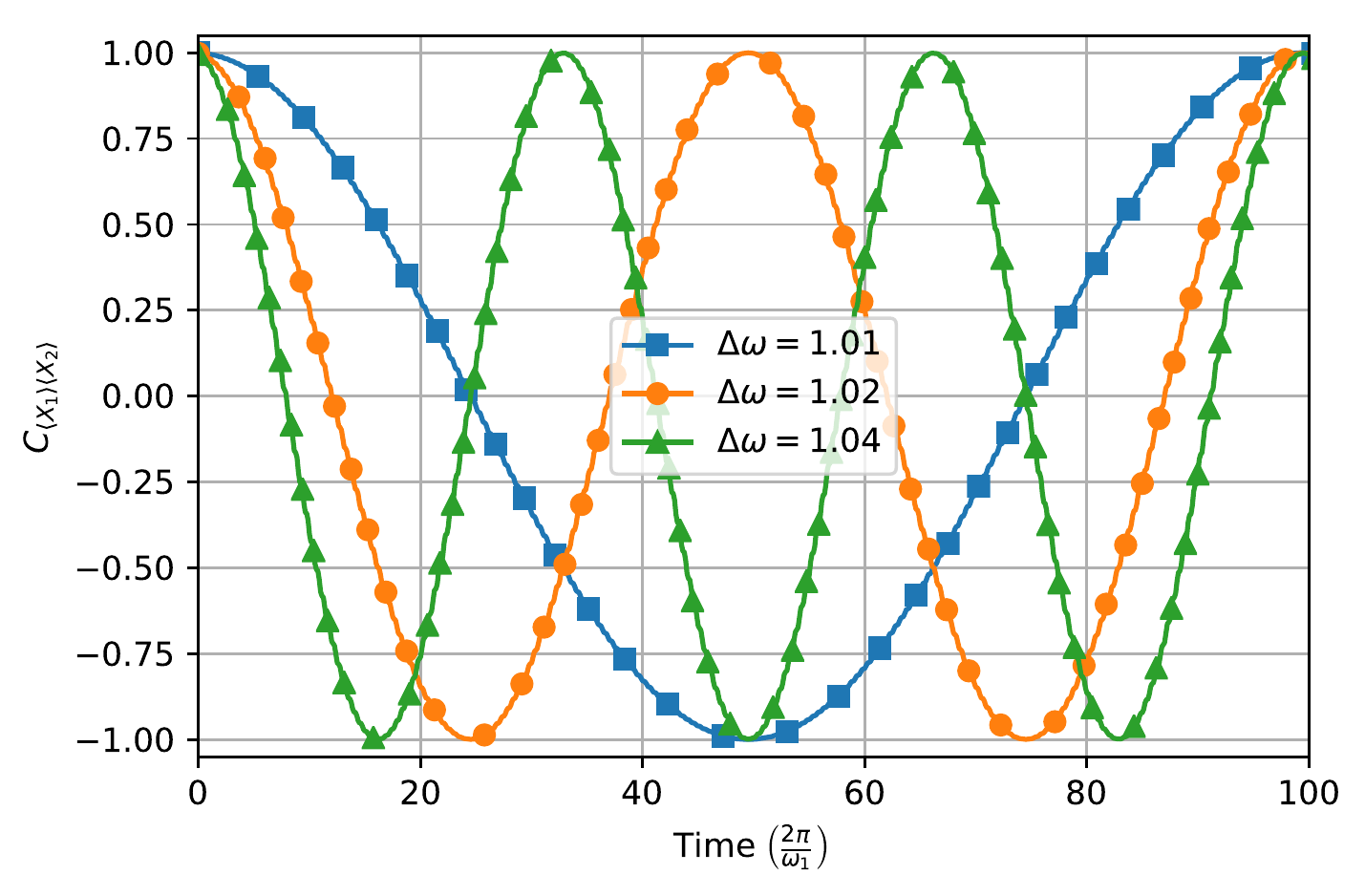}
    \caption{Value of the transient spontaneous synchronisation measure $C_{f_1,f_2}(\Delta t = 2\pi/\omega_1)$ for three sets of two uncoupled sinusoids $f_1 = \sin(\omega_1 t)$, $f_2 = \sin(\omega_2 t + \phi)$ with different frequencies $\Delta \omega = \omega_2/\omega_1$ and constant phase difference $\phi$.}
    \label{fig: characterising_C_detuning}
\end{figure}

In Fig.~\ref{fig:model_C_phase_lag}(b), the value of the synchronisation function is plotted as a function of constant phase difference $\phi$. The synchronisation measure for three different scenarios is illustrated in Fig.~\ref{fig: characterising_C_detuning}. We observe that for two waves of different frequency, the synchronisation measure $C_{f_1,f_2}\left(t|\Delta t=T_1\right)$ does not stabilise and oscillates as the phase relationship shifts over time. The frequency of oscillation is proportional to the frequency difference between the two oscillators.

These effects can be understood by analysing the form of Eq.~(\ref{eqn: synchronisation product integral}). For any two functions, Eq.~(\ref{eqn: synchronisation product integral}) has a maximum when they are identical, and therefore has a maximum value of 1. For any phase shift or frequency shift the integral of their product will be less than the square root of the product of their integrals.

This measure is used repeatedly throughout the paper as a dynamic measure of phase and allows us to connect synchronisation phase to the Hamiltonian structure and quantum correlations in novel ways. Other attempts to develop a real-time phase measure between two oscillating signals have been conducted along the lines of sliding-window discrete Fourier transform methods \cite{Orallo2015}, Hilbert transforms with data extension \cite{Shen2015} and other correlation functions \cite{Shen2016}.


\section{Emergence of synchronisation phase}\label{sec:Results}

We first consider the effect of detuning in the exciton-vibration dimer and find the emergence of a synchronisation phase in Section~\ref{sec: detuning in dimer model result}. Then, we determine the origin of the synchronisation phase in Section~\ref{sec: origin of synch phase with detuning}, which for our model is due to asymmetric vibronic interactions.

\subsection{Detuning in the bio-inspired dimer}\label{sec: detuning in dimer model result}

We define the detuning $\Delta \omega = \omega_2/\omega_1$ and choose to change the frequency of $\omega_2$ only. This allows us to fix the time window of the synchronisation measure, $C_{\langle X_1 \rangle,\langle X_2 \rangle}(t|\Delta t)$, as $\Delta t = \frac{2\pi}{\omega_1}$ whilst still probing detuning. For simplicity in the notation, from here on we denote our time-dependent measure for this time window simply $C_{\langle X_1 \rangle,\langle X_2 \rangle}(t)$.

\begin{figure}
    \centering
    \includegraphics[width=\columnwidth]{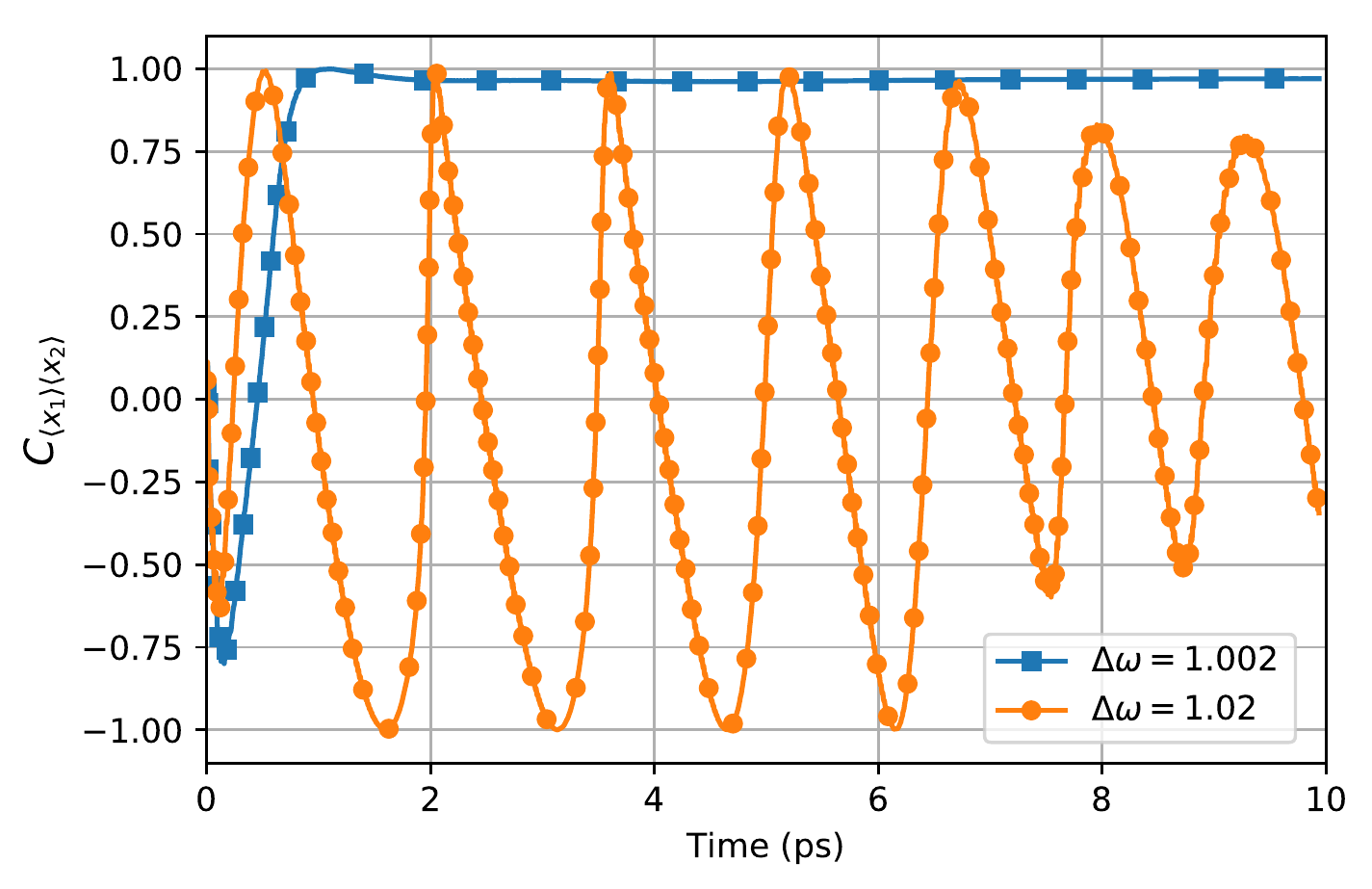}
    \caption{Transient spontaneous synchronisation measure $C_{\langle X_1\rangle,\langle X_2 \rangle}(t)$ of the expectation value of mode positions $\langle X_1\rangle,\langle X_2 \rangle$ for two regimes of detuning, with frequency difference $\Delta \omega = 1.002$ where synchronisation occurs, and $\Delta \omega = 1.02$ where it does not.}
    \label{fig: synch and no synch PE545}
\end{figure}

Using the initial state Eq.~(\ref{eqn: initial state}) and system parameters listed in Table \ref{table: numerical parameters}, we show in Fig.~\ref{fig: synch and no synch PE545} the effects on the synchronisation measure for two different regimes of detuning, $\Delta \omega = 1.002$ and $\Delta \omega = 1.02$. These are chosen to illustrate two distinct scenarios in synchronisation dynamics, namely synchronised and not synchronised respectively. Recall that here, a constant $C$ value of the measure corresponds to synchronisation.

We find that for detuning $\Delta \omega = 1.02$, the synchronisation measure oscillates, indicating that the phase between the two oscillators is continuously changing. Hence, the frequencies at which the observables $\langle X_1 \rangle$ and $\langle X_2 \rangle$ oscillate are different and have not synchronised. We have not shown here, but we also find that the phase oscillation frequencies are correlated with increased detuning. This is as expected from classical dynamics of Eq.~(\ref{eqn: classical phase equation}). The phase relationship $\Delta \phi (t)$ outside the synchronisation region would change at a rate proportional to their detuning $d$ with periodic fluctuations from $\epsilon q(\Delta \phi (t))$.

In contrast, we have synchronisation in the case of small detuning: the straight line for $\Delta \omega = 1.002$ in Fig.~\ref{fig: synch and no synch PE545} indicates that there is a constant, non-zero phase relationship between $\langle X_1 \rangle$ and $\langle X_2 \rangle$. Their frequencies have synchronised but they are not perfectly aligned in phase. This result also agrees with the predictions of Eq.~(\ref{eqn: classical phase equation}) within the synchronisation region.

\begin{figure}
    \centering
    \includegraphics[width=\columnwidth]{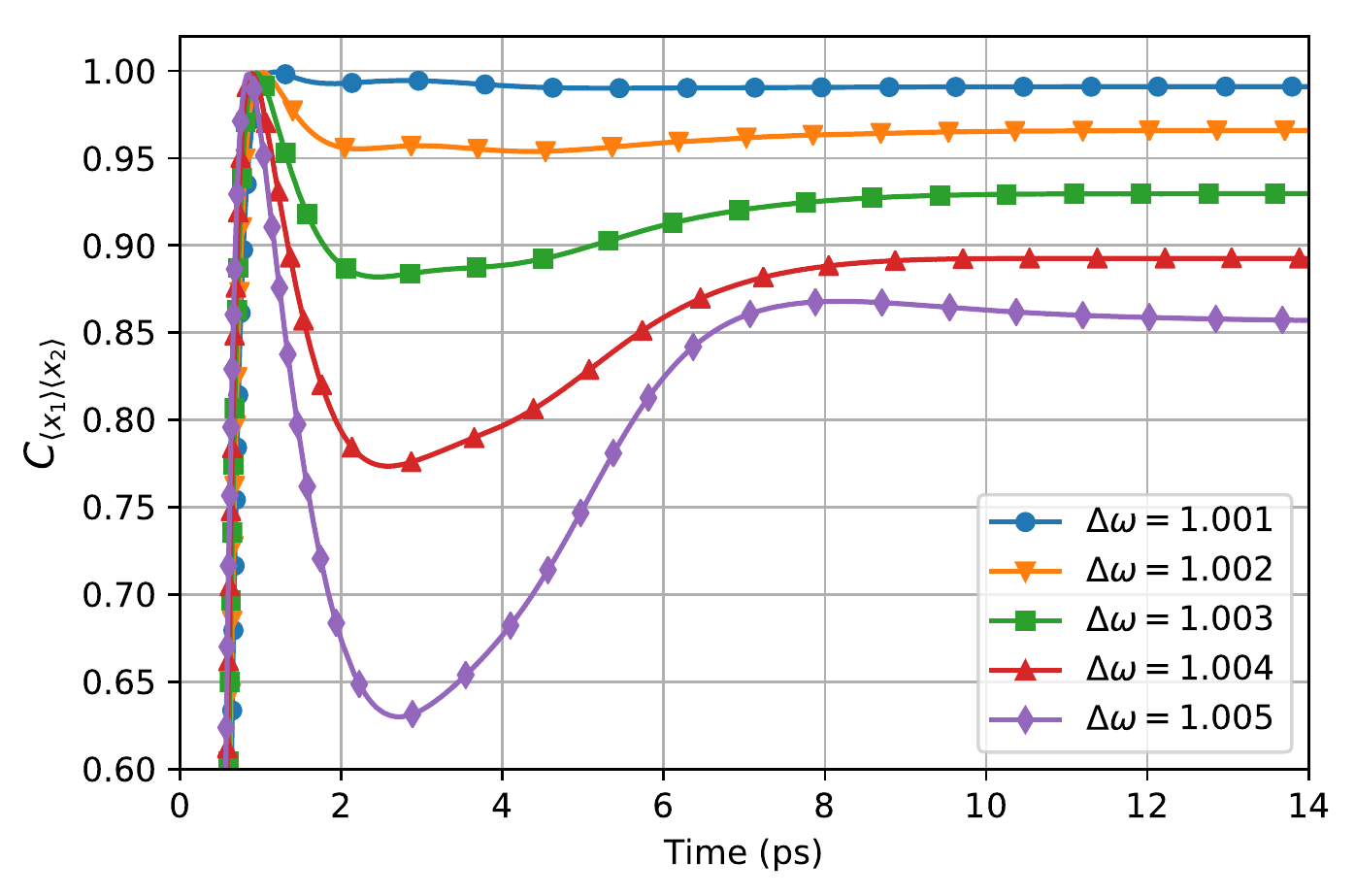}
    \caption{Transient spontaneous synchronisation measure $C_{\langle X_1\rangle,\langle X_2 \rangle}(t)$ of the expectation values of mode positions, $\langle X_1\rangle,\langle X_2 \rangle$, revealing synchronisation phase as a function of detuning, $\Delta \omega$ (listed in-figure). To highlight the long-time behaviour, the figure omits the very short-time regime.}
    \label{fig: synch with constant phase PE545}
\end{figure}

To take a closer look at this relationship we calculate the synchronisation dynamics for a range of detunings, with results given in Fig.~\ref{fig: synch with constant phase PE545}. The figure shows only the detunings for which a synchronised state is reached, i.e., having a constant long-time $C_{\langle X_1\rangle,\langle X_2\rangle}$ value. We see a clear negative relationship between the magnitude of detuning and the stable value of $C_{\langle X_1\rangle,\langle X_2\rangle}(t)$ reached, i.e., increasing detuning increases the synchronisation phase, up to a breaking point where synchronisation can no longer occur.

The overall dynamics of the synchronisation measure $C_{\langle X_1\rangle,\langle X_2\rangle}(t)$ at different detunings presented in this section are similar to what would be expected in the classical case. There exists a window of detunings within which synchronisation occurs and the time taken to reach a synchronised state increases with detuning. The quantum mechanical mechanism for this observed behaviour is explained fully in the following section.

\subsection{Origin of synchronisation phase with detuning}
\label{sec: origin of synch phase with detuning}

The origin of the relationship between the synchronisation phase and detuning can be understood by considering how the associated changes in the Hamiltonian lead to an asymmetric participation of the local modes in the collective vibronic eigenstates of the system and in the dynamics. The equations of motion for each mode's average displacement are
\begin{equation}
    \begin{split}
    \langle X_1 \rangle & = \sum_{j,k} \rho_{jk}(t) X_{1,kj}, \\
    \langle X_2 \rangle & = \sum_{j,k} \rho_{jk}(t) X_{2,kj},
    \end{split}
    \label{eqn: synch condition open}
\end{equation}
where $X_{i,kj} = \langle\psi_k|X_i|\psi_j\rangle$ with the eigenstates $|\psi_j\rangle$ given in Eq.~(\ref{eq:eigenstates_for_H}). When $\omega_1=\omega_2$, the elements $X_{i,kj}$ were restricted to either being equal or opposite upon mode exchange as thoroughly discussed in  Ref.~\cite{Siwiak-Jaszek2019}. In this situation each mode participates equally or oppositely in every vibronic coherence (c.f. Table 2 in Ref.~\cite{Siwiak-Jaszek2019}, where the values of $X_{i,kj}$ are equal or scaled by -1). As  synchronisation is the result of one specific vibronic coherence significantly out-living others, the resultant synchronisation phase was restricted to either $0$ (when $C_{\langle X_1\rangle,\langle X_2 \rangle}=1$) or $\pi$ (when $C_{\langle X_1\rangle,\langle X_2 \rangle}=-1$) depending on which coherence dominated. For the undetuned situation in our bio-inspired system, mode displacements are found to be perfectly synchronised and phase-matched in the long-time regime. Later on in Section \ref{sec: militello et al case study} we discuss how this reflects a collective normal mode being effectively decoupled from the electronic dynamics.

With unequal frequencies $\omega_1 \neq \omega_2$, the mode exchange symmetry is broken leading to an asymmetric participation of each mode both in the vibronic eigenstates and in the dynamics. In the far detuning extreme, one mode may be so far off-resonance with the system energy scales that it does not participate in system dynamics at all and synchronisation cannot occur. In the smaller detuning regime, the eigenstates structure do not restrict $X_{1,jk}$ and $X_{2,jk}$ to be symmetric and anti-symmetric but instead they have a range of amplitudes. The time dependent parts of Eq.~(\ref{eqn: synch condition open}) are identical for $\langle X_1 \rangle$ and $\langle X_2 \rangle$ but the weights of each oscillating component changes according to $X_{1,jk}$ and $X_{2,jk}$. Within the region of synchronisation, these amplitude differences are small and manifest as a constant phase difference between the oscillations of each expected value $\langle X_i \rangle$. This can be seen by expressing each amplitude as $X_{i,kj} = \exp(\kappa_{i,kj})$ where $\kappa$ is in general a complex number thereby contributing a phase to the signal. Outside the region of synchronisation, these differences are large, the signals are composed of different frequencies and have an unstable phase difference.

\section{Quantum correlations \label{sec:quantum correlations}}

Given that the exciton-vibration dimer operates in the quantum regime, a question of interest is: how much of its behaviour is uniquely quantum?
The quantitative relationships between spontaneous quantum synchronisation and quantum discord, a measure of purely quantum correlation, have been explored in a range of quantum synchronisation settings \cite{Ameri2015,Benedetti2016,Galve2010,Giorgi2012,Giorgi2013,Manzano2013,Manzano2013a,Witthaut2017,Zhu2015}. For example, Ref. \cite{Giorgi2012} find that the emergence of spontaneous synchronisation is correlated with the preservation of quantum discord. Spontaneous synchronisation can also be correlated with the generation of entanglement from an initially unentangled state \cite{Benedetti2016}, and the quantum mutual information has been proposed as a measure of synchronisation that is capable of being used in both deep quantum and semi-classical regimes \cite{Ameri2015}.

In Section~\ref{sec: quantum correlation measures}, we introduce measures of quantum correlations. In Section~\ref{sec: QC results in dimer}, we investigate the dynamics of quantum correlations between the spontaneously synchronising subsystems, and find further evidence for the connection between synchronisation and quantum correlations.
Specifically, we reveal that the synchronisation phase indicates a change in magnitude of quantum discord between the synchronising subsystems. This suggests that our adapted synchronisation measure can be used to quantify a purely quantum feature.

\subsection{Quantum correlation measures}
\label{sec: quantum correlation measures}

The quantum mutual information $I(A:B)$ is a measure of the total correlations between two subsystems $A$ and $B$ of a bipartite quantum system $AB$ and is defined as:
\begin{equation}
    I(A:B) = S(\rho_A) + S(\rho_B) - S(\rho_{AB}),
    \label{eqn: mutual information}
\end{equation}
where $S(\rho) = -\tr\left[\rho \log \rho\right]$ is the von Neumann entropy and density matrices $\rho_A = \tr_B\left[\rho_{AB}\right]$, $\rho_B = \tr_A\left[\rho_{AB}\right]$ are subsystems of $\rho_{AB}$. This shared information can be decomposed into classical correlations and quantum correlations. The classical correlations are equivalently the difference in von Neumann entropy of a subsystem before and after a measurement is acted on the other subsystem:
\begin{equation}
    J(B|A) = \max_{A^\dag_i A_i} \left\{S(\rho_B) - \sum_i p_i S(\rho^i_B) \right\},
    \label{eqn: classical information}
\end{equation}
where
\begin{equation}
    \rho_B^i = \tr_A\left[A^\dag_i A_i \rho_{AB} \right] / p_i,
\end{equation}
is the residual state of $B$ after measurement of $A^\dag_i A_i$ (positive operator valued measurements) on subsystem $A$ and $p_i = \tr_{AB}\left[A^\dag_i A_i \rho_{AB} \right]$ is the probability of this outcome. Numerically, the measurements $A^\dag_i A_i$ are generated randomly until the sum of Eq.~(\ref{eqn: classical information}) satisfactorily converges on its maximum. Note that this equation would be different for the classical correlations from subsystem $A$ to $B$, which we would label $J(A|B)$.

The remaining portion of the mutual information that is not classical must be quantum, i.e. the quantum discord $D(B|A)$  \cite{Henderson2001, Ollivier2002}:
\begin{equation}
    D(B|A) = I(A:B) - J(B|A).
    \label{eqn: quantum discord}
\end{equation}

\subsection{Spontaneous synchronisation and quantum correlations}
\label{sec: QC results in dimer}

\begin{figure}
    \centering
    \includegraphics[width=\columnwidth]{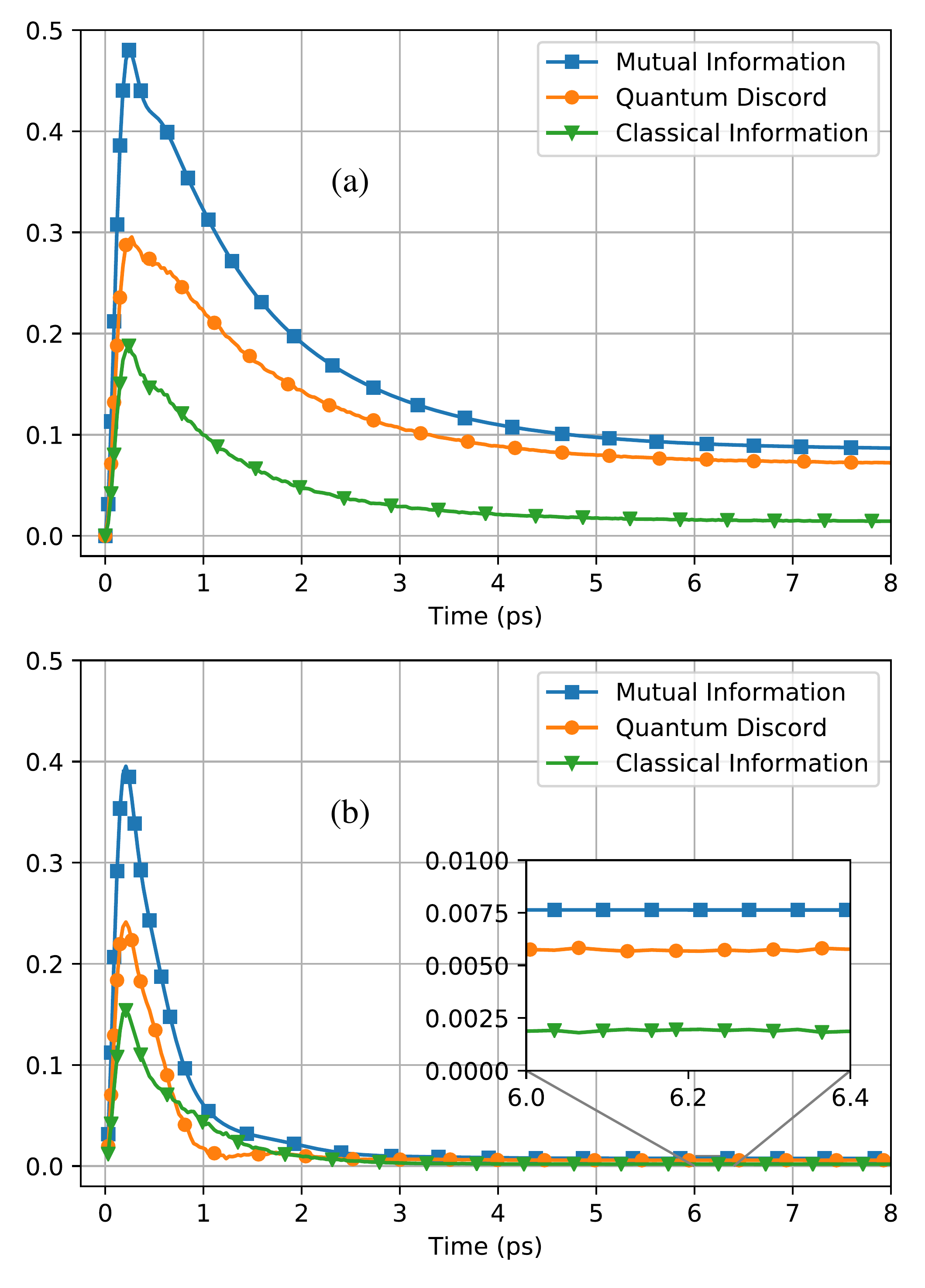}\hfill
    \caption{Dynamics of quantum mutual information, quantum discord and classical information between mode subsystems in the exciton-vibration dimer model with PE545 parameters (Table \ref{table: numerical parameters}) and initial state Eq.~(\ref{eqn: initial state}) for two detuning scenarios: (a) $\Delta \omega = 1.002$ for which transient spontaneous synchronisation is achieved and (b) $\Delta \omega = 1.02$ for which it is not.}
    \label{fig: PE545 QCs for synch and non-synch}
\end{figure}

Given the structure of our dimer model in Section \ref{sec:exciton-vibration dimer}, the density matrices for the subsystems are accessible.  Hence, we can use  Eqs.~(\ref{eqn: classical information}), (\ref{eqn: mutual information}) and (\ref{eqn: quantum discord}) to calculate the quantum correlations between mode subspaces. Intuitively we would expect some mutual information between modes to be maintained in a synchronised state as, if they were completely uncorrelated, then they should oscillate at independent frequencies and phases. From the work in Ref. \cite{Siwiak-Jaszek2019}, we know that in the systems considered, synchronisation requires vibronic eigenstates with sufficient participation from both modes. We postulate that these quantum correlations would not persist in the long-time limit if synchronisation is not achieved. The detuning scenarios introduced in this paper provide the ideal regime to test this.

Fig.~\ref{fig: PE545 QCs for synch and non-synch} shows numerical calculations of the mutual information, quantum discord and classical information between the two intramolecular modes with PE545 parameters and for two different scenarios: Fig.~\ref{fig: PE545 QCs for synch and non-synch}(a) considers the case with detuning $\Delta \omega = 1.002$ in which synchronisation occurs, while Fig.~\ref{fig: PE545 QCs for synch and non-synch}(b) considers $\Delta \omega = 1.02$ in which synchronisation does not occur.

Two time-regimes emerge. Firstly, we notice a sharp increase in all correlations from uncorrelated initial state at $\unit[0]{ps}$ to a peak at around $\unit[0.2]{ps}$. Note that the correlations do not originate from the initial state---instead, they must be generated by the coherent interactions in our system. The peak at $\unit[0.2]{ps}$ coincides with the coherent excitation transfer period that is characteristic of the dynamics in these bio-inspired vibronic dimers (c.f. Fig.~8c from Ref.~\cite{Siwiak-Jaszek2019} and Fig.~10b from Ref.~\cite{Kolli2012}). The excitation transfer mechanism involves transitions between vibronic eigenstates which involve both modes, therefore it is unsurprising that the quantum correlations between the modes also peak at the same time.

Secondly, we note the decay in correlations from $\unit[0.4]{ps}$ onward. This behaviour is due to the decay of coherent dynamics and the dominance of incoherent processes. In the synchronising case of Fig.~\ref{fig: PE545 QCs for synch and non-synch}(a) we see that the modes remain significantly correlated in the long time limit whereas in Fig.~\ref{fig: PE545 QCs for synch and non-synch}(b) we see that correlations decay rapidly to a much lower value. This clear correlation between synchronisation and the preservation of quantum correlations is in agreement with previous findings.

Interestingly, we find that when the systems spontaneously synchronise, the majority fraction of the mutual information consists of quantum discord at every instant in time. In contrast, if the systems do not synchronise, we see a time period in which classical information is greater than quantum discord as it can been noticed in the $\unit[1]{ps}$ to $\unit[2]{ps}$ time interval of Fig.~\ref{fig: PE545 QCs for synch and non-synch}(b). This leads us to hypothesise a novel dynamical relationship between transient spontaneous synchronisation and quantum correlations: for spontaneous quantum synchronisation to emerge, the quantum discord must be greater than classical information at all times, whereas for non-synchronising cases there may exist time intervals where the greater fraction of correlations are classical.

\begin{figure}
    \centering
    \includegraphics[width=\columnwidth]{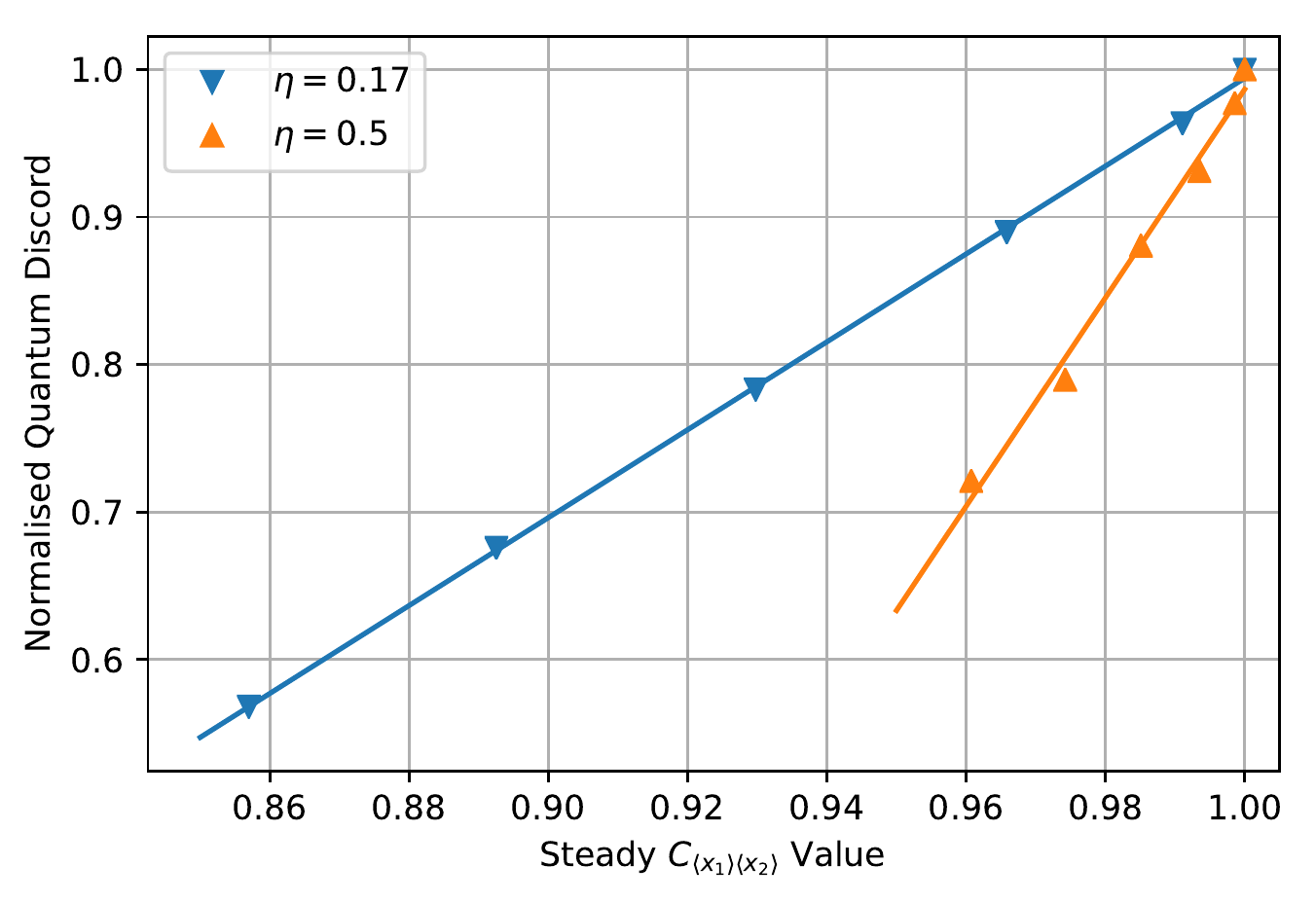}
    \caption{Long-time stable value of quantum discord between mode subsystems plotted against long-time stable value of transient spontaneous synchronisation measure $C_{\langle X_1\rangle,\langle X_2 \rangle}(t)$ of expectation value of mode positions. Each data point corresponds to a value of detuning and quantum discord is normalised to $1$ for zero detuning. Two degrees of excitonic delicalisation given by $\eta=2V/|\Delta e|$ are considered: (i) $\eta = 0.17$ corresponding to parameters as in Table \ref{table: numerical parameters} and (ii) adjusted electronic values such that $\eta = 0.5$ but all remaining parameters are the same as in  in Table \ref{table: numerical parameters}. Linear regressions are plotted using standard methods to emphasise the relationship.}
    \label{fig: discord connected to phase PE545}
\end{figure}

Now consider the in-between detuning regime for which synchronisation phase is also achieved. In Fig.~\ref{fig: discord connected to phase PE545}, we plot the long-time value of quantum discord (normalised to the discord at no detuning) and compare with the long-time constant synchronisation measure $C_{\langle X_1\rangle,\langle X_2\rangle}$. We consider two different exciton delocalisation regimes of our dimer model characterised by different values of $\eta=2V/|\Delta e|$ \cite{Siwiak-Jaszek2019}: (i) the long blue line is for the parameters and detunings from Fig.~\ref{fig: synch with constant phase PE545} giving $\eta=0.17$ and (ii) the short orange line for a slightly more delocalised regime where $\Delta e$ and $V$ are adjusted such that $\eta = 0.5$ but all the remaining parameters are as in Table 1. Each point corresponds to a different detuning that exhibits a synchronisation phase (cf.  Sec. \ref{sec: detuning in dimer model result}).
Hence, for our bio-inspired system, a larger synchronisation phase upon detuning, corresponding to a stable value $0<C_{\langle X_1\rangle,\langle X_2\rangle}<1$, indicates weaker quantum correlations between the modes. This shows that our synchronisation measure is able to capture the change in quantum discord between subsystems as a function of frequency detuning.
In the next section we explore how general our observations are when considering other forms of interaction between two-level systems and Harmonic oscillators.

\section{The case of Militello et al.'s model}
\label{sec: militello et al case study}

In order to understand how the emergence of the phase synchronisation in the presence of detuning is linked to the structure of Hamiltonian and to the dissipative dynamics the system experiences, we consider the study by Militello et al. \cite{Militello2017}. The authors demonstrate synchronisation phase in a open quantum system that has comparable features our bio-inspired model but that exhibits a non-zero synchronisation phase in the absence of detuning. We discuss the physical origin of their synchronisation phase and demonstrate the agreement with our approach and predictions.

\subsection{Synchronisation phase without detuning}

The model considered by Militello et al. consists of a two-level system with ground and excited stated denoted as $|e_1\rangle$ and  $|e_2\rangle$ respectively, and interacting with two quantum harmonic oscillators whose associated lowering operators are $b_1$ and  $b_2$. The total Hamiltonian has a Jaynes-Cummings form:
\begin{equation}
    \begin{split}
        H^{\text{Militello}} = & + e_1 |e_1\rangle\langle e_1| + e_2 |e_2\rangle\langle e_2| \\
        & + \omega_1 b_1^\dag b_1 +\omega_2 b_2^\dag b_2 \\
        & + g_1\left(e^{i\phi_1}b_1 + e^{-i\phi_1}b_1^\dag\right)\sigma_x \\
        &+ g_2\left(e^{i\phi_2}b_2 + 
        e^{-i\phi_2}b_2^\dag\right)\sigma_x,
    \end{split}
    \label{eqn: militello Hamiltonian}
\end{equation}
where $\sigma_x = |e_1\rangle\langle e_2| + |e_2\rangle\langle e_1|$ and $\phi_{1}$ and $\phi_{2}$  are phases in the range $0\leq\phi\leq\pi$. Each two-level system is also coupled to a zero-temperature reservoir modelled by a Markovian master equation of the form
\begin{equation}
    \dot{\rho}(t) = -i[H^{\text{Militello}},\rho(t)] + D_{\sigma_-}[\rho(t)],
\end{equation}
where $D_{\sigma_-}$ is a Lindblad superoperator of the form given in Eq.~(\ref{eq:lindblad}) with transition operator $\sigma_- = |e_1\rangle\langle e_2|$ and rate $\Gamma_{\sigma_-}$. The modes do not experience any direct dissipation.

The authors find that when synchronisation occurs, there is a constant phase difference between the two mode displacement that is determined by the parameters $\phi_1$ and $\phi_2$ from $H^{\text{Militello}}$ and obeys the relation
\begin{equation}
    \phi_{S} = \pi - (\phi_1-\phi_2).
    \label{eqn: militello synch phase}
\end{equation}
Note that their approximated synchronisation phase has no dependence on detuning and is a function of the variables $\phi_1$ and $\phi_2$ only. 
Militello et al. derive this relation by approximating the effects of dissipation on an initial coherent state of the modes. Since synchronisation phase for Militello et al.'s model can emerge for zero-detuning and symmetric coupling strength, i.e.  $\omega_1=\omega_2$ and $g_1=g_2$, it is important to understand the physical origin such a phase in this situation and how it is subsequently altered as detuning is introduced.

\begin{figure}
    \centering
    \includegraphics[width=\columnwidth]{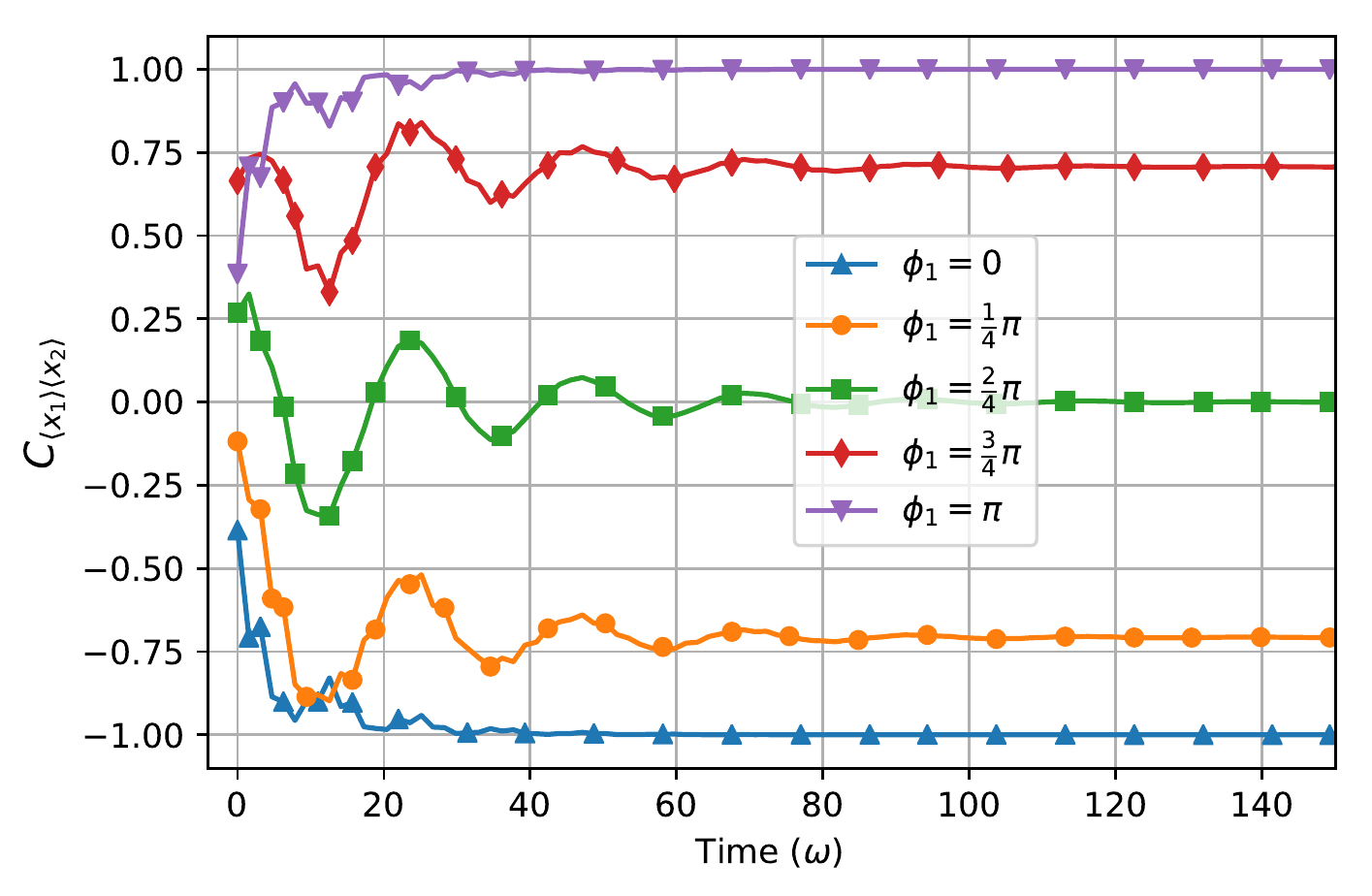}
    \caption{Synchronisation measure $C_{\langle X_1\rangle,\langle X_2 \rangle}(t)$ of expectation value of mode positions for a different values of (oscillation difference) $\phi_1$ as shown in Eq.~(\ref{eqn: militello interactions phi_1}) for Militello et al.'s model  \cite{Militello2017}. Initial state: $\rho_0 = |e_1\rangle\langle e_1| \otimes \rho_{\text{vacuum}} \otimes \rho_{\text{coherent}}$. Parameters $\Delta e = e_2-e_1$, $\omega = \Delta e$, $g = \Delta e$, $\Gamma_{\sigma_-} = 0.2 \Delta e$.}
    \label{fig: militello synch phase due to phi}
\end{figure}

In order to gain this understanding we focus on the structure of  the interaction part of the Hamiltonian in Eq.~(\ref{eqn: militello Hamiltonian}) by fixing $\phi_2 = 0$ and letting $\phi_1$ control the specific form of the interaction between the modes and the two-level subsystem:

\begin{subequations}
    \begin{align}
        H_I^{\text{Militello}} (\phi_1 = 0) & =  g \sigma_x \left(X_2 + X_1\right) \\
        H_I^{\text{Militello}} \big(\phi_1 = \frac{1}{4} \pi \big) & =  g \sigma_x \left(X_2 - \frac{X_1}{\sqrt{2}} - \frac{P_1}{\sqrt{2}}\right) \\
        H_I^{\text{Militello}} \big(\phi_1 = \frac{2}{4} \pi\big) & =  g \sigma_x \left(X_2 + P_1\right) \\
        H_I^{\text{Militello}} \big(\phi_1 = \frac{3}{4} \pi\big) & =  g \sigma_x \left(X_2 + \frac{X_1}{\sqrt{2}} + \frac{P_1}{\sqrt{2}}\right) \\
        H_I^{\text{Militello}} (\phi_1 = \pi) & =  g \sigma_x (X_2 - X_1),
    \end{align}
    \label{eqn: militello interactions phi_1}
\end{subequations}
where $P_1$ is the dimensionless momentum operator for mode 1. Evolution in each of these scenarios results in synchronisation of the mode observables $\langle X_i \rangle$ with different phase differences $\phi_S$. In Fig.~\ref{fig: militello synch phase due to phi}, we consider Militello et al.'s model and plot the correlation $C_{\langle X_1\rangle,\langle X_2 \rangle}(t)$ for the five example interaction Hamiltonians in the set of Eqs.~(\ref{eqn: militello interactions phi_1}). Using the relationship between 
$C_{\langle X_1\rangle,\langle X_2\rangle} $ and the synchronisation phase presented 
in Fig.~\ref{fig:model_C_phase_lag}(b), we can see that our predictions for the synchronisation phase agree in all cases with the predictions of Militello et al.

For every case in Eqs.~(\ref{eqn: militello interactions phi_1}) there are effective two normal collective modes. As dissipation acts directly only on the two-level sub-system, the dynamics of the oscillators depend critically on which collective observables are involved in the interaction $H_I^{\text{Militello}}$. In the case of Eq.~(\ref{eqn: militello interactions phi_1}a), the sum of the mode position operators, also known as the collective `centre-of-mass' mode $X_+ =(X_1+X_2)$, is directly coupled to the two-level subsystem. In this case, the collective `relative-displacement' mode $X_- =( X_1-X_2)$ is fully decoupled and is therefore free from dissipation. When evolving from an initial coherent state that contains some amplitude in both of these collective modes, the centre-of-mass motion decays rapidly while the relative-displacement remains. As we show in Fig.~\ref{fig: militello synch phase due to phi}, survival of this collective mode in the long-time regime implies perfectly anti-correlated motions ($C_{\langle X_1\rangle,\langle X_2\rangle}~=~-1$) indicating a constant $\pi$ phase between $\langle X_1 \rangle$ and $\langle X_2 \rangle$ in agreement with Eq.~(\ref{eqn: militello synch phase}) i.e. $\phi_S = \pi - (0-0) = \pi$.

Similar analysis holds for the case of Eq.~(\ref{eqn: militello interactions phi_1}e) where the collective center-of-mass motion decouples from the interaction with the two-level system and  long-time synchronisation of displacements is positive ($C_{\langle X_1\rangle,\langle X_2\rangle}=1$) with perfect phase match i.e. $\phi_S=0$. This case is comparable with our exciton-dimer model with zero-detuning and symmetric coupling, i.e. $\omega_1=\omega_2$ and $g_1=g_2=g$, as it can be noted by expressing the interaction part of Eq.~(\ref{System Hamiltonian}) in terms of effective Pauli matrices for the excitonic system \cite{OReilly2014}:
\begin{equation}
    H_I (\text{dimer}) = g (\cos2\theta\sigma_z + \sin2\theta\sigma_x)(X_2 - X_1), 
    \label{eq:HI_ours}
\end{equation}
where $\theta=\frac{1}{2}\arctan (2V/|\Delta e|)$. For the parameters of our bio-inspired dimer, $\theta$ is small and the interaction with $\sigma_z$ dominates in contrast with $H_{I}^{\text{Militello}}$ where the interaction couples $\sigma_x$. The specificity of the $\sigma_x$ or $\sigma_z$ interaction manifests itself in different short-time synchronisation dynamics but the stable long-time synchronisation phase is dominated by the effectively decoupled collective mode as discussed above. We can then conclude that for each in-between value of $\phi_1$ and the corresponding Hamiltonians, i.e. Eqs.~(\ref{eqn: militello interactions phi_1}b)-(\ref{eqn: militello interactions phi_1}d), there exists a collective mode that is decoupled both from the two-level system and from the collective mode entering $H_{I}^{\text{Militello}}$, and which dominates the long-time synchronisation dynamics as it is indeed discussed by Militello et al. (cf. Eqs.~(7) and (8) in \cite{Militello2017}).

In summary, the syncnronisation phase predicted by Ref. \cite{Militello2017} for zero-detuning is entirely determined by the relative phases of the collective mode that decouples fully both from the two-level system and from the collective mode entering $H_{I}^{\text{Militello}}$. When frequency detuning is introduced, the local mode exchange symmetry is broken leading to their asymmetric participation in  dynamics and to collective modes which cannot be decoupled from each other (cf. Eq.~(3a) in Ref.~\cite{Militello2017}.) In the presence of a small detuning, synchronisation is still determined by a collective mode that decouples from the two-level system but that remains weakly coupled to the collective mode entering the interaction and therefore undergoes indirect dissipation. Thus, a shift in the synchronisation phase given by Eq.~(\ref{eqn: militello synch phase}) shall be expected as a function of detuning as we show in the next subsection.

\subsection{Detuning in Militello et al.'s model}

\begin{figure}
    \centering
    \includegraphics[width=\columnwidth]{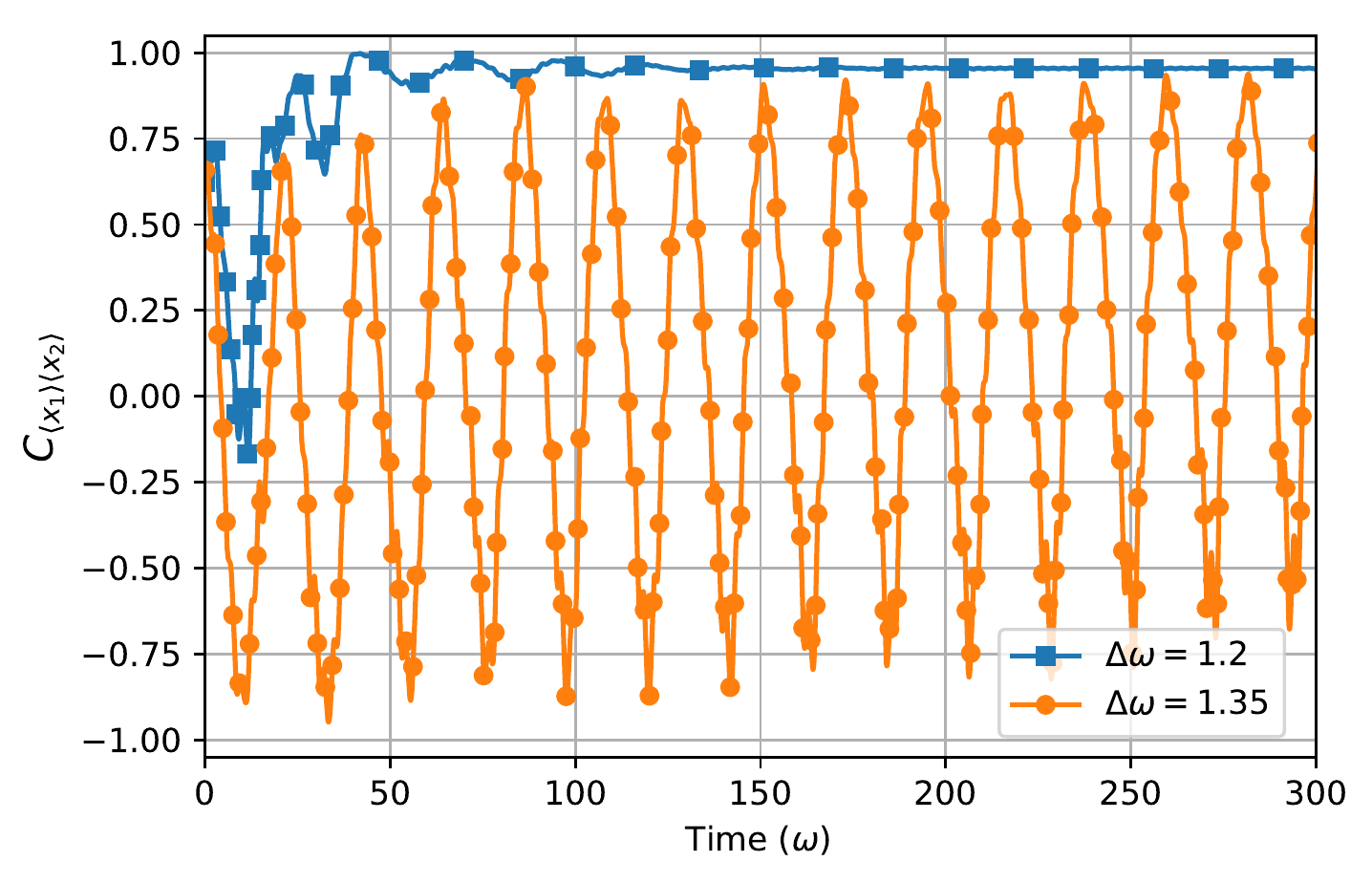}
    \caption{Transient spontaneous synchronisation measure $C_{\langle X_1\rangle,\langle X_2 \rangle}(t)$ of expectation value of mode positions for two regimes of detuning in Militello et al.'s model  \cite{Militello2017}: detuning $\Delta\omega = 1.2$ where synchronisation occurs and $\Delta \omega = 1.35$ where it does not. Initial state: $\rho_0 = |e_1\rangle\langle e_1| \otimes \rho_{\text{vacuum}} \otimes \rho_{\text{coherent}}$. Parameters $\Delta e = e_2-e_1$, $\omega = \Delta e$, $g = \Delta e$, $\Gamma_{\sigma_-} = 0.2 \Delta e$.}
    \label{fig: synch and no synch militello}
\end{figure}

We now investigate how detuning further influences the synchronisation phase in Militelo et al.'s model. To understand this we consider the Hamiltonian in Eq.~(\ref{eqn: militello Hamiltonian}) with $\phi_1=\pi$ and $\phi_2=0$ and detune the mode frequencies such that  $\Delta \omega = \omega_2/\omega_1>1$. The results are reported in Fig.~\ref{fig: synch and no synch militello} and they strongly resemble the behaviour shown in Fig.~\ref{fig: synch and no synch PE545} whereby a small detuning i.e. $\Delta \omega =1.2$ renders a synchronised state with a long-time constant value of $C_{\langle X_1\rangle,\langle X_2\rangle}$ slightly less than 1 thereby signalling a non-zero phase synchronisation. As expected, a larger detuning prevents synchronisation.

We expect a similar behaviour for all values of $\phi_1$ and $\phi_2$ when including detuning, with the understanding that the undetuned situation sets the level from which the shift in the synchronisation phase should be measured. The results in Fig.~\ref{fig: synch and no synch militello} also support the understanding that the synchronisation phase in the presence of detuning accounts for the asymmetric participation of the modes in the eigenstates and dynamics such that no collective degree of freedom is fully isolated from dissipation. Our analysis then shows that the observations in Section \ref{sec: detuning in dimer model result} are not exclusive to the exciton-vibration dimer and apply to a variety of spontaneously synchronising quantum systems.

\subsection{Quantum correlations in Militello et al.'s model}
\label{sec: QCs in militello et al}
In the following we measure the quantum correlations between undetuned modes spontaneously synchronising with a constant phase as produced by Militello et al.'s  Hamiltonian Eq.~(\ref{eqn: militello Hamiltonian}) with identical mode frequencies and interaction strengths. 

\begin{figure}[t]
    
    \includegraphics[width=\columnwidth]{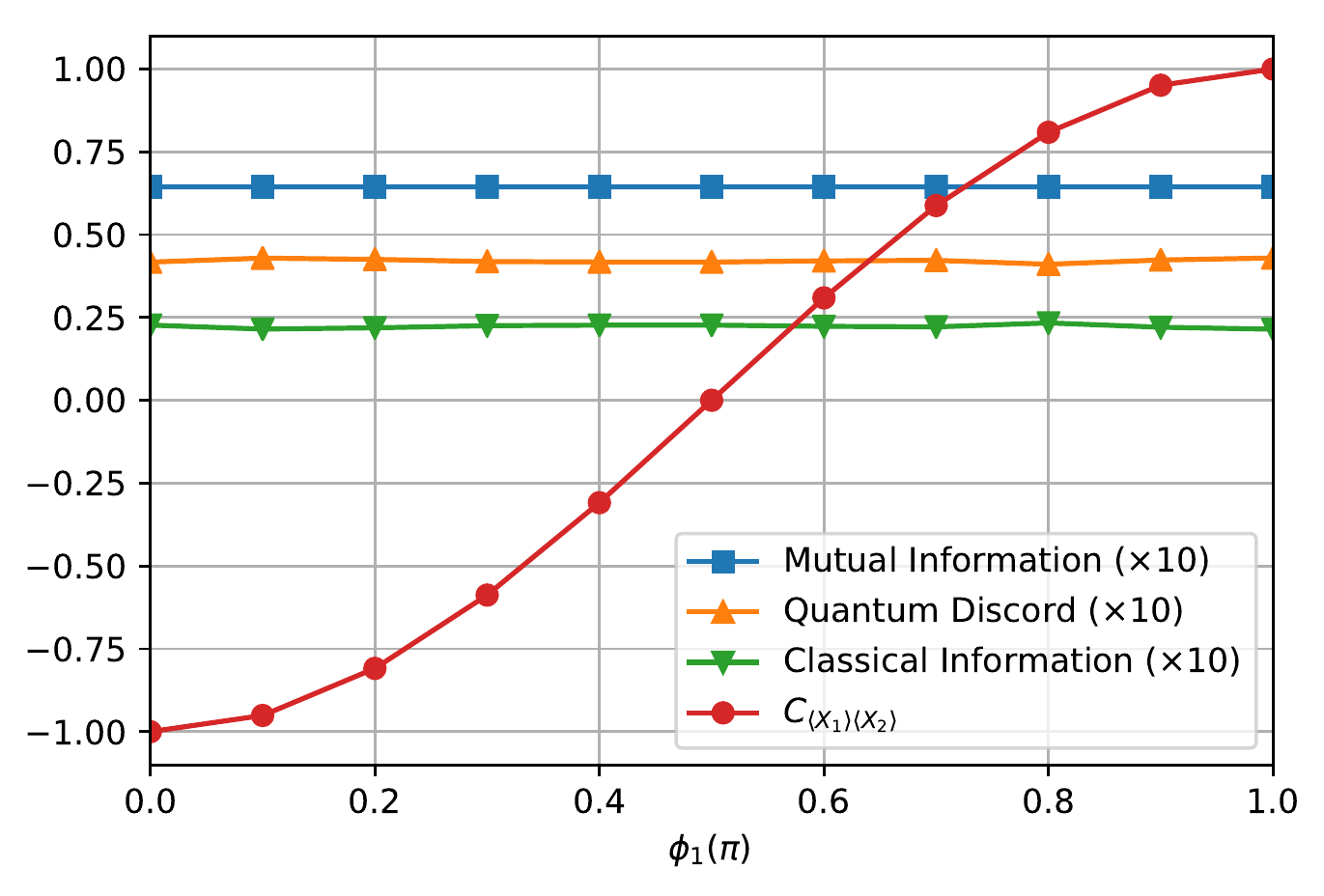}
    \caption{Long-time limit stable values of quantum mutual information, quantum discord and classical information between mode subsystems as functions of $\phi_1$ for Militello et al.'s interaction Hamiltonian given in Eq.~(\ref{eqn: militello Hamiltonian}) with $\phi_2=0$. Values for these functions have been scaled by 10.  Also shown is the long-time limit stable value of the synchronisation function $C_{\langle X_1\rangle,\langle X_2\rangle}$ as function of $\phi_1$.}
    \label{fig: QCs_MLT_dphi}
\end{figure}

In Fig.~\ref{fig: QCs_MLT_dphi} we plot the long-time stable values of quantum mutual information, classical information and quantum discord for a range of $\phi_1$ alongside the corresponding long-time stable value  $C_{\langle X_1\rangle,\langle X_2\rangle}$ which quantifies the synchronisation phase. We find that the long time correlations are unchanged by the phase $\phi_1$ introduced in the interaction Hamiltonian. This is expected since changes in $\phi_1$ do not change the coupling strengths or individual frequencies of the local modes, and hence shall not affect the long-time correlations.
The synchronisation $C_{\langle X_1\rangle,\langle X_2\rangle}(t)$ dynamics presented as a function of $\phi_1$ is exactly what we would expect from two sine functions with a constant phase shift between them, as can be seen in the characterisation of our synchronisation function presented in Fig.~\ref{fig:model_C_phase_lag}(b). We do expect that for a fixed set of $\phi_1$ and $\phi_2$ parameters, the long-time quantum correlations and synchronisation phase will change for different detunings as reported in Fig. \ref{fig: discord connected to phase PE545}.

\begin{figure}[!ht]
   \centering
    \includegraphics[width=\columnwidth]{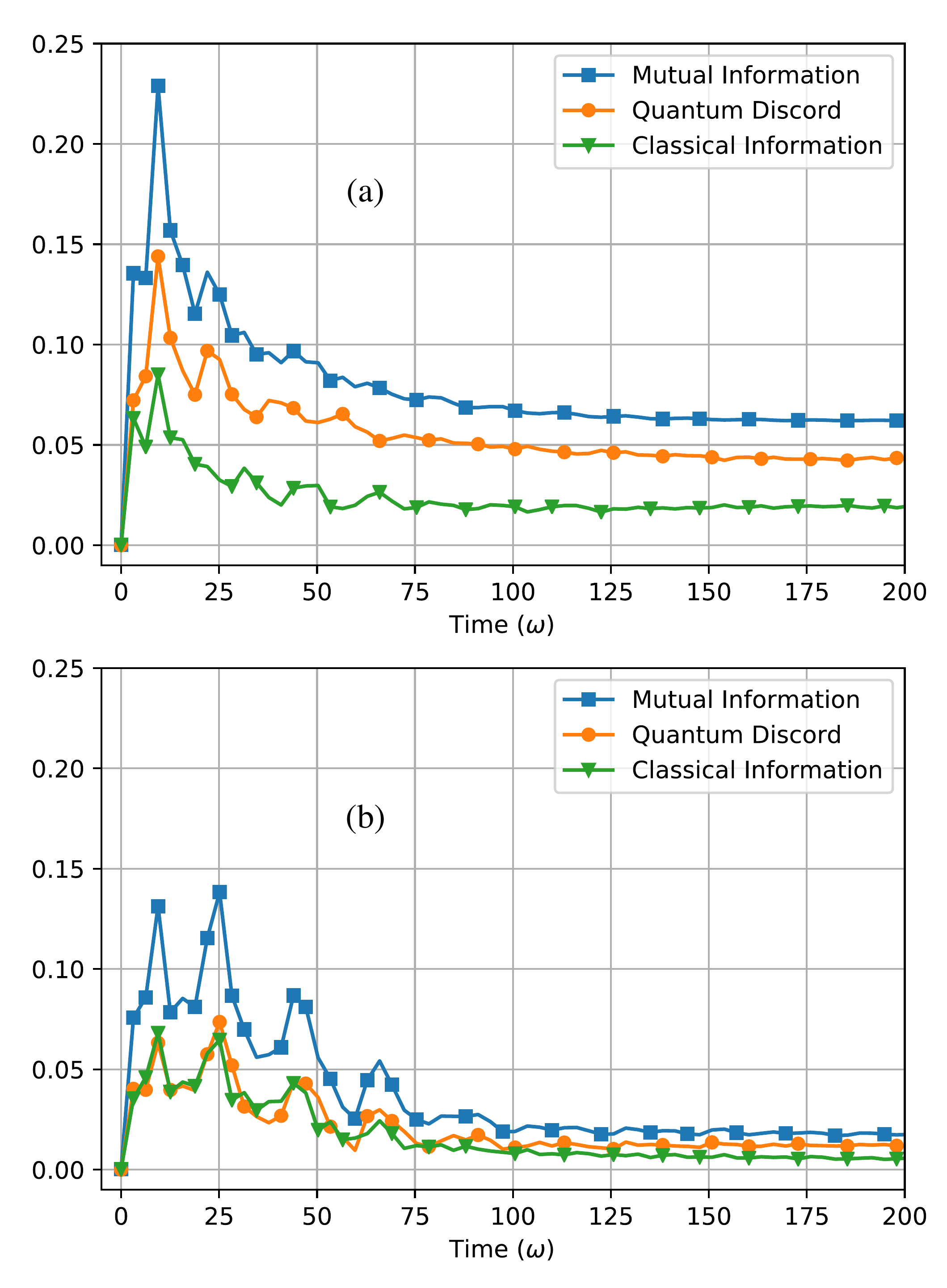}
    \caption{Dynamics of quantum mutual information, quantum discord and classical information between mode subsystems in Militello et al.'s model \cite{Militello2017} for two detuning scenarios: (a) where synchronisation is achieved $\Delta \omega = 1$ and (b) where it is not $\Delta \omega = 1.35$. Initial state: $\rho_0 = |e_1\rangle\langle e_1| \otimes \rho_{\text{vacuum}} \otimes \rho_{\text{coherent}}$. Parameters $\Delta e = e_2-e_1$, $\omega = \Delta e$, $g = \Delta e$, $\Gamma_{\sigma_-} = 0.2 \Delta e$.}
    \label{fig: MLT QCs for synch and non-synch}
\end{figure}

Finally, in Fig.~\ref{fig: MLT QCs for synch and non-synch}, we investigate the dynamics of  quantum and classical correlations in Militello et al's model with $\phi_1=\pi$ and $\phi_2=0$ and for synchronising and non-synchronising regimes. We find the same qualitative relationship as in the exciton-vibration dimer model with detuning reported in Fig.~\ref{fig: PE545 QCs for synch and non-synch}. This supports the generality of the hypothesis introduced in Sec.~\ref{sec: QC results in dimer} that for spontaneous synchronisation to emerge the dynamics must be such that quantum discord is larger than the classical information at all times, whereas in non-synchronising systems classical information can outweigh quantum discord for non-zero time periods.

\section{\label{sec:discussion_and_conclusion} Conclusion }

In this paper we have defined a synchronisation measure that effectively quantifies the synchronisation phase between oscillating signals and employ this measure to investigate the dynamics of transient spontaneous synchronisation as a function of detuning in a bio-inspired vibronic dimer subjected to  Markovian dissipation. We also explore the quantum and classical nature of the information shared between synchronising subsystems and how this information dynamics is captured by our proposed synchronisation measure.

For the bio-inspired quantum system of interest, where local vibrational modes assist electronic energy transfer, we determined the detune regime in which spontaneous synchronisation can occur with a constant non-zero synchronisation phase. Transient spontaneous synchronisation in this system reflects the fact that a collective motion undergoes very weak dissipation. In the case of equal oscillator frequencies, the sychronised state in the long-time regime renders zero phase difference between the synchronising subsystems \cite{Siwiak-Jaszek2019}. We showed that upon introducing detuning, the synchronisation phase is shifted. The origin of this phase shift can be traced back to the asymmetric participation of the vibrations in the joint vibronic eigenstates and in the dynamics. In a normal mode picture this means that collective normal motions are now coupled to each other such that no collective mode is entirely decoupled from dissipative processes thereby affecting the phase at which local vibrations syncrhonise. Through the study of Militello et al.'s model \cite{Militello2017}, we showed that the mechanisms affecting the shift in the synchronisation phase under detuned conditions applies to a variety of synchronising open quantum systems.

We then investigated the relationship between transient spontaneous synchronisation and quantum correlations, in both our exciton-vibronic dimer and in the model of Militello et al. \cite{Militello2017}. We found that if synchronisation occurs, then the shared information between synchronising subsystems is primarily quantum discord at all times, whereas in the non-synchronising cases classical information may some times be the larger fraction. We also found that as a function of the detuning, the quantum discord between synchronising subsystems decreases linearly as the synchronisation measure decreases, that is, the shift in the synchornisation phase upon detuning implies that quantum correlations between the subsystems persist but are diminished. Our results then suggest that our measure is capable of capturing information about a purely quantum property of synchronising subsystems, and that the measure can indicate the persistent presence and change in quantum discord.

We have considered the simplest bio-inspired quantum scenario capturing key features present in a variety of natural photosynthetic light-harvesting complexes to illustrate that quantum synchronisation analysis provides a insightful route for understanding truly  quantum phenomena in such systems.  Thus, our work opens up a promising avenue to investigate non-trivial quantum phenomena in a variety of complex biomolecular and chemical systems.

\bibliography{Mendeley}
\bibliographystyle{apsrev4-1}
\end{document}